\documentclass[12pt,letterpaper]{article} 
\usepackage[verbose,margin=1in]{geometry}
\usepackage{cite}
\usepackage{epsfig}
\usepackage[margin=1cm]{caption}
\usepackage{color}
\usepackage{amsmath}
\usepackage{amssymb}
\usepackage{enumitem}
\usepackage{booktabs}
\usepackage{authblk}
\usepackage[firstpage]{draftwatermark}

\SetWatermarkText{\includegraphics[angle=-45,height=11in]{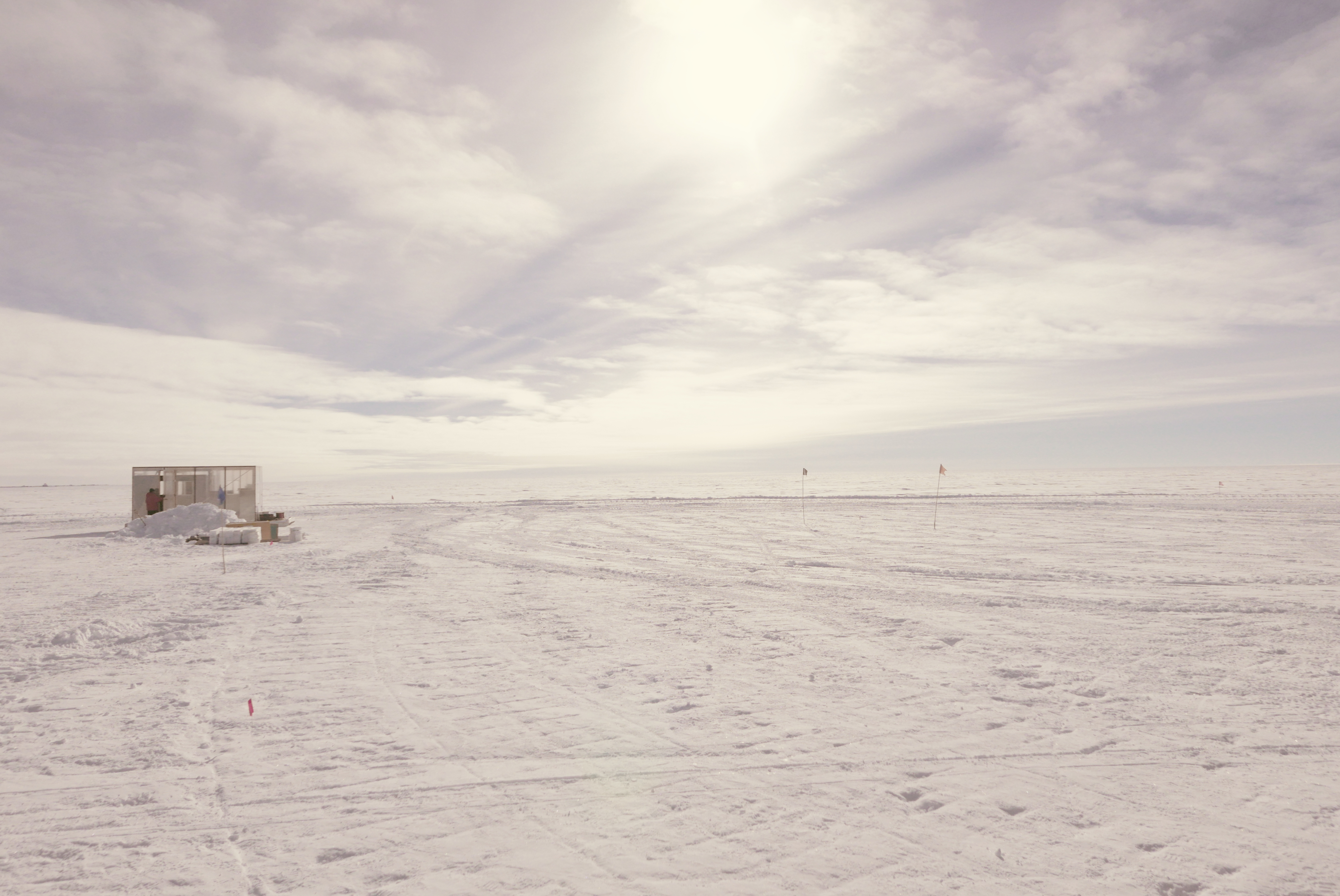}}

\definecolor{titleblue}{rgb}{0.0,0.5,0.9}
\usepackage[small]{titlesec}

\usepackage[implicit=true, debug=false, 
plainpages=false, colorlinks=true, 
linkcolor=black, urlcolor=black, citecolor=black, filecolor=black, 
hyperindex=true]{hyperref}

\begin{document}
\setlist[itemize]{noitemsep, topsep=0pt}

\title{\textcolor{titleblue}{\bf The Next-Generation Radio Neutrino Observatory -- Multi-Messenger Neutrino Astrophysics at Extreme Energies}\vspace{1in}}
\date{}
\author[1]{J. A. Aguilar}
\author[3]{P. Allison}
\author[4]{S. Archambault}
\author[3]{J. J. Beatty}
\author[6]{D. Z. Besson}
\author[7]{\\O. Botner}
\author[2]{S. Buitink}
\author[8]{P. Chen}
\author[3]{B. A. Clark}
\author[3]{A. Connolly\thanks{\href{mailto:connolly@physics.osu.edu}{\textcolor{blue}{connolly@physics.osu.edu}}}}
\author[9]{C. Deaconu}
\author[2]{\\S. de Kockere}
\author[5]{M. A. DuVernois}
\author[2]{N. van Eijndhoven}
\author[10]{C. Finley}
\author[11]{D. Garcia}
\author[7]{\\A. Hallgren}
\author[5]{F. Halzen}
\author[12]{J. Hanson}
\author[5]{K. Hanson}
\author[7]{C. P{\'e}rez de los Heros}
\author[13]{\\K. D. Hoffman}
\author[5]{B. Hokanson-Fasig}
\author[9]{K. Hughes}
\author[10]{K. Hultqvist}
\author[15]{A. Ishihara} 
\author[5]{A. Karle}
\author[5]{J. L. Kelley}
\author[15]{S. R. Klein}
\author[11]{M. Kowalski}
\author[16]{I. Kravchenko}
\author[6] {U. A. Latif}
\author[8]{T. C. Liu} 
\author[5]{M.-Y. Lu} 
\author[14]{K. Mase}
\author[5]{R. Morse}
\author[8]{J. Nam}
\author[11]{A. Nelles}
\author[9]{E. Oberla}
\author[17]{\\C. Pfendner}
\author[18]{Y. Pan}
\author[11]{I. Plaisier}
\author[3]{S. Prohira}
\author[15]{S. Robertson}
\author[3]{J. Rolla}
\author[19]{\\D. Ryckbosch}
\author[18]{F. G. Schr{\"o}der}
\author[18]{D. Seckel}
\author[16]{A. Shultz}
\author[9]{D. Smith}
\author[9]{D. Southall}
\author[10]{\\E. O'Sullivan}
\author[1]{S. Toscano}
\author[3]{J. Torres-Espinosa}
\author[7]{E. Unger}
\author[9]{\\A. G. Vieregg}
\author[2]{K. de Vries}
\author[8]{S.-H. Wang}
\author[11]{C. Welling}
\author[20]{S. A. Wissel}
\author[14]{S. Yoshida}

\affil[1]{Université Libre de Bruxelles, Science Faculty CP230, B-1050 Brussels, Belgium}
\affil[2]{Vrije Universiteit Brussel, Dienst ELEM, B-1050 Brussels, Belgium}
\affil[3]{Dept. of Physics and Center for Cosmology and Astro-Particle Physics, Ohio State University, Columbus, OH 43210, USA}
\affil[4]{Dept. of Physics and Institute for Global Prominent Research, Chiba University, Chiba 263-8522, Japan}
\affil[5]{Dept. of Physics and Wisconsin IceCube Particle Astrophysics Center, University of Wisconsin, Madison, WI 53706, USA}
\affil[6]{Dept. of Physics and Astronomy, University of Kansas, Lawrence, KS 66045, USA}
\affil[7]{Dept. of Physics and Astronomy, Uppsala University, Box 516, S-75120 Uppsala, Sweden}
\affil[8]{Dept. of Physics, Grad. Inst. of Astrophys., \& Leung Center for Cosmology and Particle Astrophysics, National Taiwan Univ., Taipei, Taiwan}
\affil[9]{Dept. of Physics and Kavli Institute for Cosmological Physics, The University of Chicago, Chicago, IL, USA}
\affil[10]{Oskar Klein Centre and Dept. of Physics, Stockholm University, SE-10691 Stockholm, Sweden}
\affil[11]{DESY, D-15738 Zeuthen, Germany}
\affil[12]{Dept. of Physics and Astronomy, Whittier College, Whittier, CA, USA}
\affil[13]{Dept. of Physics, University of Maryland, College Park, MD 20742, USA}
\affil[14]{Dept. of Physics and Institute for Global Prominent Research, Chiba University, Chiba 263-8522, Japan}
\affil[15]{Lawrence Berkeley National Laboratory, Berkeley, CA 94720, USA}
\affil[16]{Dept. of Physics and Astronomy, University of Nebraska-Lincoln, Lincoln, NE, USA}
\affil[17]{Dept. of Physics, Otterbein University, Westerville, OH 43081, USA}
\affil[18]{Bartol Research Institute and Dept. of Physics and Astronomy, University of Delaware, Newark, DE 19716, USA}
\affil[19]{Dept. of Physics and Astronomy, University of Gent, B-9000 Gent, Belgium}
\affil[20]{Dept. of Physics, California Polytechnic State University, San Luis Obispo, CA, USA}

\maketitle 
\thispagestyle{empty}

\newpage
\abstract{
 RNO is the mid-scale discovery instrument designed to make the first observation of neutrinos from the cosmos at extreme energies, with sensitivity well beyond current instrument capabilities.
This new observatory will be the largest
ground-based neutrino telescope to date, 
enabling the measurement of neutrinos above $10^{16}$~eV, determining the nature of the astrophysical neutrino flux that has been measured by IceCube at higher energies, similarly extending the reach of multi-messenger astrophysics to the highest energies, and enabling investigations of fundamental physics at energies unreachable by particle accelerators on Earth.}\\

{\bf Thematic activities:}  Multi-messenger Astronomy and Astrophysics, Cosmology and Fundamental Physics

\pagenumbering{roman} 

\newpage

\newpage
\pagenumbering{arabic}

\section{Introduction}

The field of radio detection of high-energy cosmic particles has been growing rapidly over the past decade.  This proposed Radio Neutrino Observatory (RNO) capitalizes on the success and combined expertise of the ARA~\cite{Allison:2011wk},
ARIANNA~\cite{Barwick:2015ica,Barwick:2014pca}, 
ANITA~\cite{anita3} 
and RICE~\cite{rice06} 
experiments, for a design that will deliver world-leading measurements of the high-energy neutrino flux, with the pointing and energy resolution required for multi-messenger astronomy, aligned with the Astro2020 Science whitepapers~\cite{astro2020_nuastro,astro2020_MMA}.  The portion of RNO deployed in the initial two years will serve
as a pathfinder for the radio component of IceCube-Gen2.

RNO consists of 61 stations located on a grid with a spacing of 1.25~km.  Each station has a surface component and a deep component (60~m below the surface of the ice)
    that together enable the detection and detailed reconstruction of neutrino events. 
RNO was designed to be located at the South Pole, taking advantage of the infrastructure
available with the South Pole station and IceCube and the suitability of the 
radio-transparent Antarctic ice sheet.  Conceptually it could be located 
in Greenland where earlier studies established that the ice sheet is suitable for 
radio detection of neutrinos. Both sites are in consideration at the time of submission. 
Infrastructure and logistics considerations may be the driving factor 
for final site selection.

\section{The Radio Neutrino Observatory (RNO)}

With the discovery of a diffuse flux of astrophysical neutrinos~\cite{IceCube2013a,IceCube2014a,IceCube2015a,IceCube2016a,8yearnumuflux} and the identification of a multi-messenger source candidate~\cite{IceCube2018a, IceCube2018b}, the success of IceCube has established neutrinos as a powerful messenger in the exploration of the high-energy universe.  RNO will extend multi-messenger neutrino astronomy to energies
above 10~PeV. 

RNO is designed around a broad multi-messenger astrophysics program to be an instrument that measures of order ten neutrinos at the highest energies, possibly including the first discovery. 
At its lowest energies, the RNO detector will overlap in sensitivity with IceCube, expecting $10-25$ astrophysical neutrino events in five years if the measured spectral index extends unbroken to higher energies.  
The RNO collaboration will develop the techniques required to rapidly produce and respond to alerts of astrophysical transients.

\subsection{A Pathfinder for IceCube-Gen2 Radio}
Building on ARA and ARIANNA experience, the initial
  RNO stations 
 that will be deployed in the first 
two years of construction (approximately twenty stations spanning $\sim 50$~km$^2$)
will
serve as a `pathfinder' program to ensure the relevant R\&D for Gen2 is completed. 
RNO will enable the determination of the optimal station depth, the relative fractions and arrangements of surface and deep antennas, and the mode of power distribution that is the most feasible for the IceCube-Gen2 radio array, envisioned at 500~km$^2$ area.  With RNO there will be continuity in the radio effort during the time leading up the IceCube-Gen2 construction. Studies with RNO will enable further development of simulations of radio arrays to be designed for and ready to handle an array of the size envisioned for IceCube-Gen2.
Coordination between the RNO and IceCube-Gen2 collaborations
and the interplay of the timelines and logistics
of the two projects is in progress.

\section{RNO Science: the Highest Energy Neutrinos}
Neutrinos are unique messengers.  They point back to their sources and can reach us from the most distant corners of the universe because they travel undeflected by magnetic fields and unimpeded by interactions with matter or radiation.  Unlike $\gamma$-rays, which can be explained by inverse Compton scattering, the observation of high-energy neutrinos from these objects provides incontrovertible evidence for cosmic-ray acceleration, since both neutrinos and $\gamma$-rays are produced when cosmic rays interact with ambient photons or matter
 within their source.  
Resolving the sources of cosmic rays and the acceleration mechanisms will require a comprehensive multi-messenger program involving observations of cosmic rays, $\gamma$-rays, and neutrinos across many decades of energy.

\subsection{Multi-Messenger Astrophysics }

\label{sect:multmessenger}

\begin{figure}[t]
    \centering
    \includegraphics[width=5.5in]{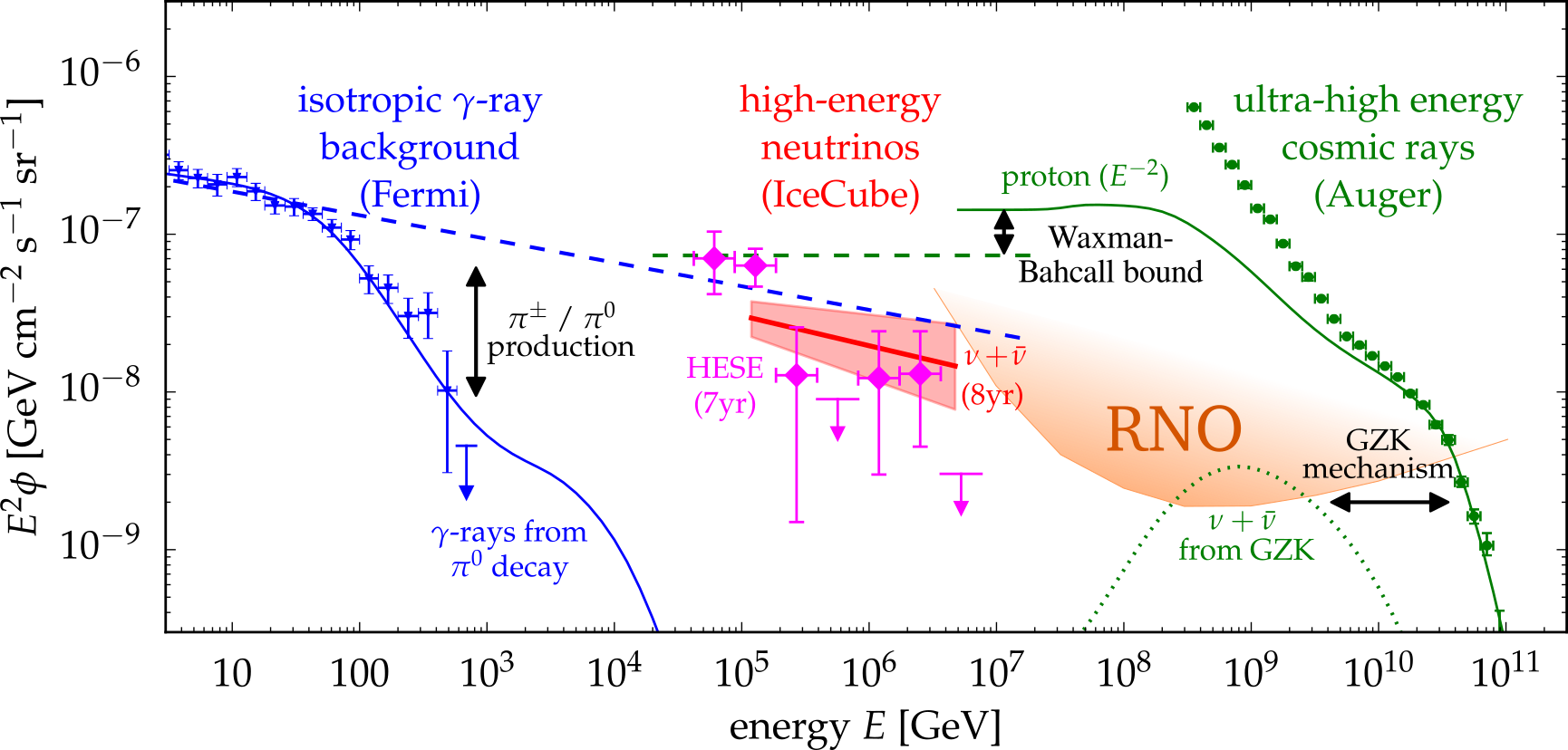}

    \caption{ \label{fig:mm_figure} The fluxes of high-energy $\gamma$-rays, neutrinos, and cosmic rays are interrelated by various processes. RNO will target both the astrophysical and cosmogenic neutrino fluxes. Modified (to show the all-flavor neutrino flux and RNO sensitivity) from~\cite{ahlers_mm_figure_source}.
    }
   
\end{figure}

In the last five years, neutrinos have delivered on their promise to provide a key piece of this astronomical puzzle with the discovery of a diffuse flux of astrophysical neutrinos~\cite{IceCube2013a,IceCube2014a,IceCube2015a,IceCube2016a}. 
IceCube has measured a spectrum of neutrinos to energies up to 10\,PeV --- the highest-energy neutrinos ever observed.  Beyond $\sim 10$~PeV, IceCube is simply too small to observe the steeply falling neutrino flux.

Figure \ref{fig:mm_figure} compares the neutrino flux measured by IceCube in the muon and cascade channels with the diffuse flux of $\gamma$-rays measured by Fermi  and the cosmic-ray spectrum  measured by Auger.
The RNO region of sensitivity sits in a crucial region where one would
observe a higher energy component of the measured astrophysical flux, as well
as neutrinos of cosmogenic origin.
The spectra from the three types of
messengers display tantalizingly similar energy densities, suggesting a common origin.  

This coincidence is even more intriguing in light of the announcement in July 2018 of the first coincident observation of a source (the blazar TXS 0506+056) flaring simultaneously in gamma-rays as well as in neutrinos~\cite{IceCube2018a, IceCube2018b}.  This was the first multi-messenger observation triggered by a high-energy neutrino, demonstrating the capability to send real time alerts and established the field as a driving force in multi-messenger astronomy.

\subsubsection{Astrophysical Neutrino Detection Rate}
\begin{figure}[t]
\hspace{6mm} \includegraphics[width=0.55\textwidth]{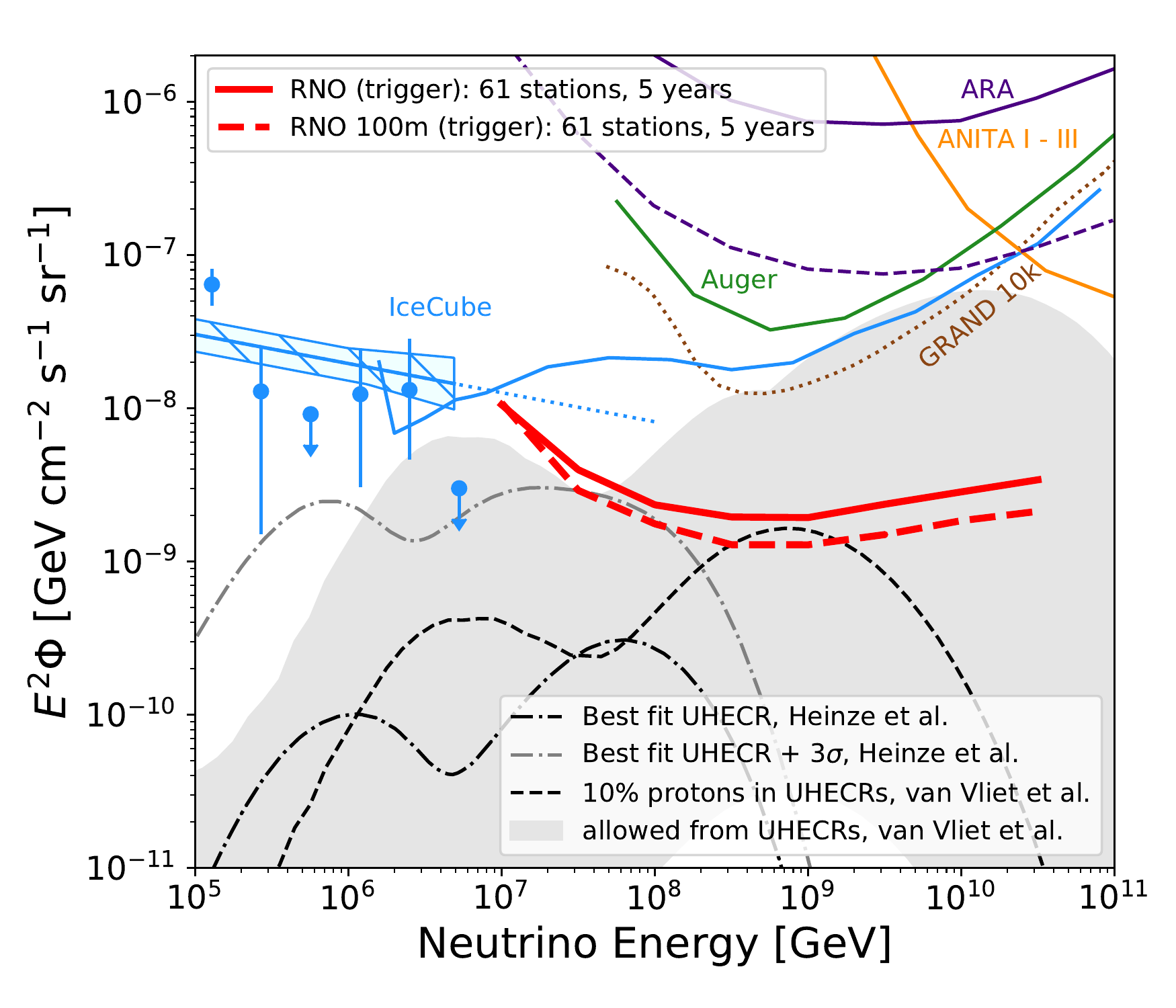}
\raisebox{2mm}{
\includegraphics[width=0.32\textwidth]{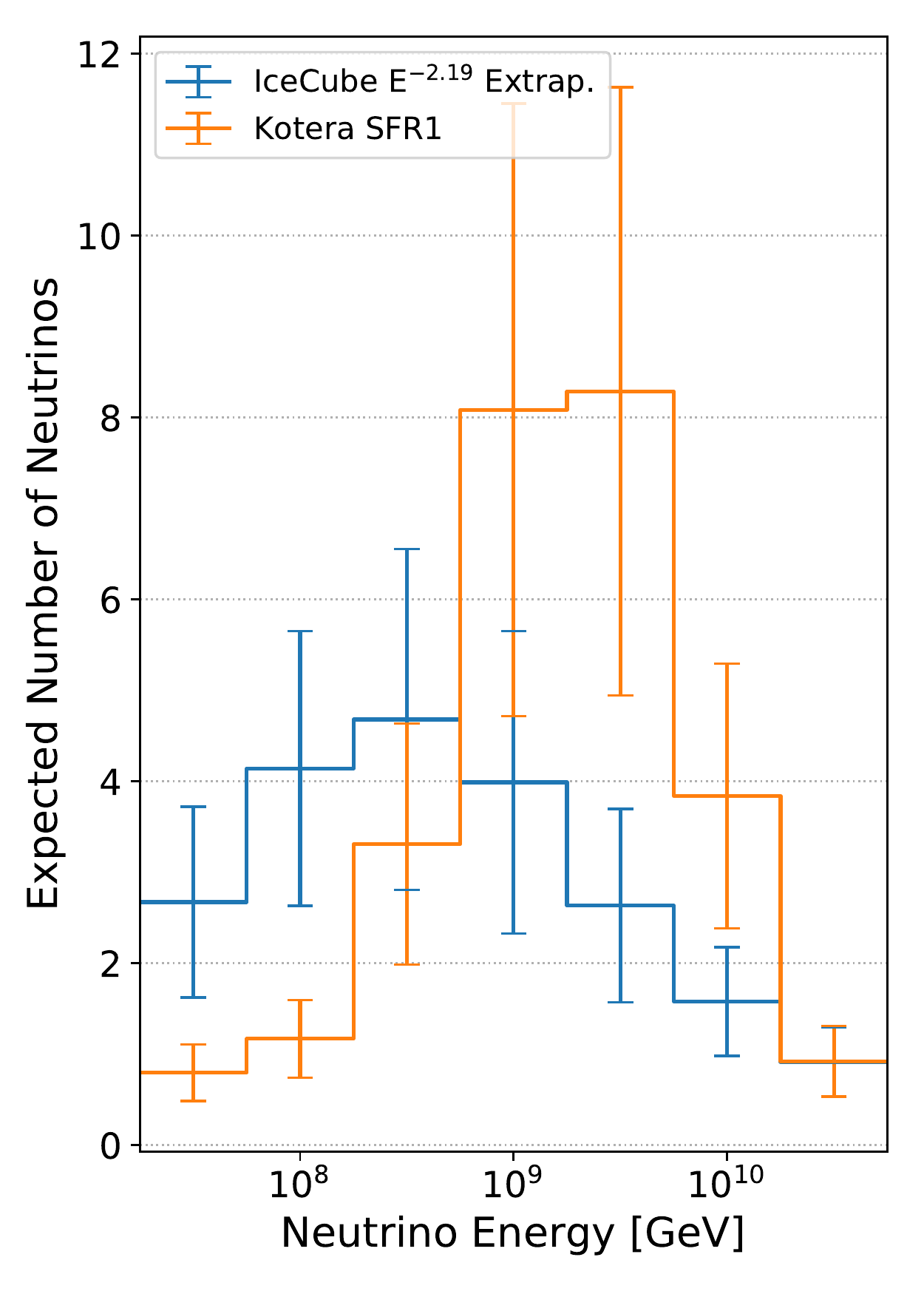}
}
\caption{Left: The sensitivity of RNO compared with current experimental results and models. All limits are in one-decade energy binning. Shown are limits from the next-gen simulation. We refer to Table~\ref{tab:event_rates} and Figure~\ref{fig:yearly_neutrinos} for event rates for different source models. For the diffuse cosmogenic neutrino flux, we show two models~\cite{Heinze,vanVliet}.  Right: The expected number of neutrinos vs. energy for the IceCube E$^{-2.19}$ extrapolated flux~\cite{8yearnumuflux} and the Kotera~et~al.~SFR1 flux~\cite{Kotera2011}. 
} 
\label{fig:limits}
\end{figure}

\begin{figure}
    \centering
    \includegraphics[width=0.39\textwidth]{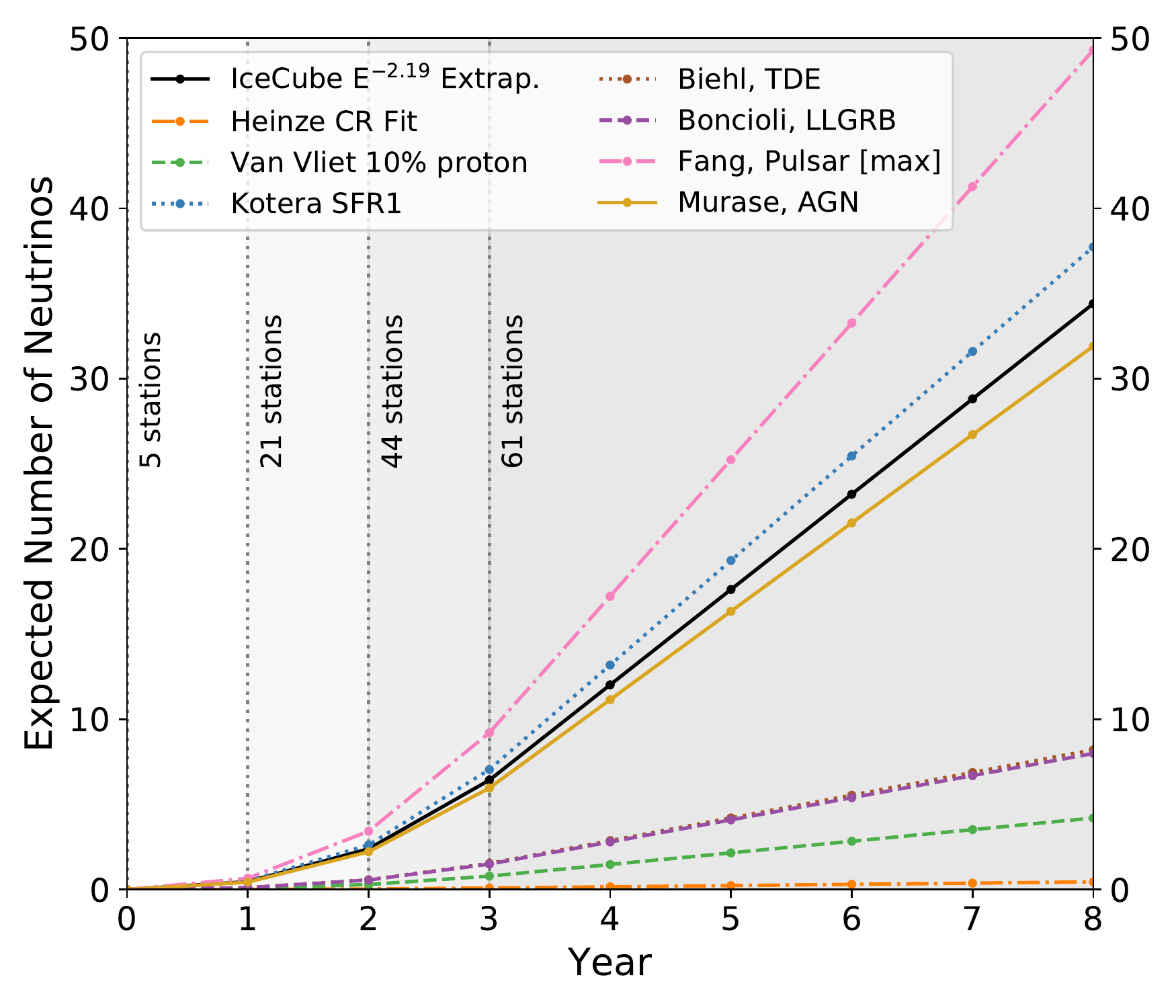}
    \raisebox{0.25in}{
    \includegraphics[width=0.47\textwidth]{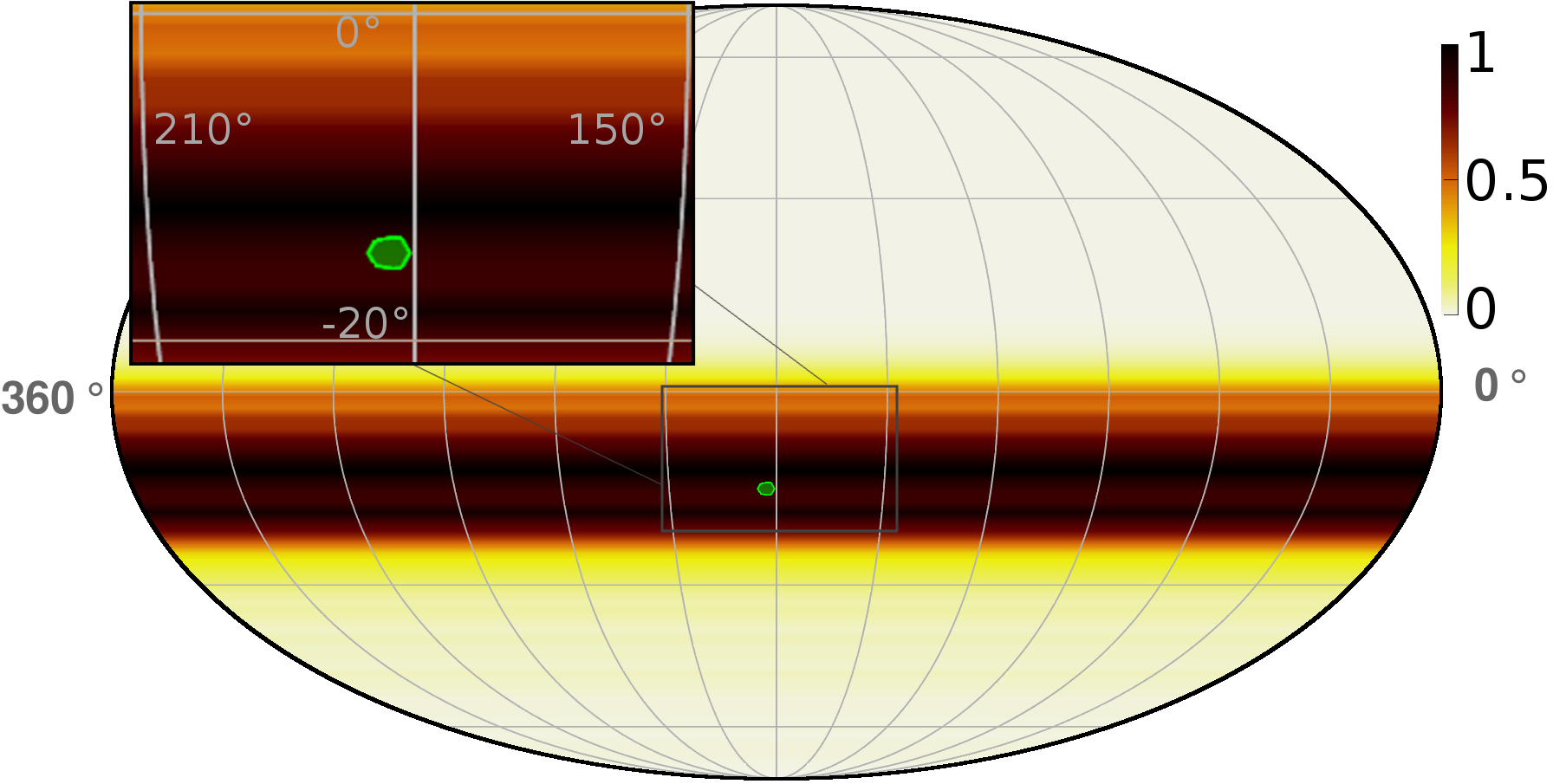}}
    \caption{(Left) The number of expected neutrinos observed as a function of time. We use an $E^{-2.19}$ flux from IceCube~\cite{8yearnumuflux} and a variety of astrophysical and cosmogenic models~\cite{Heinze, vanVliet, Kotera2011, Biehl, Boncioli, FangPulsar, MuraseAGN}. 
    The 'Hi' prediction in Table~\ref{tab:event_rates} corresponds to the
    expected number of neutrinos from five post-deployment years in this figure. (Right) The expected 90\% containment for an RNO neutrino event in equatorial coordinates in green, overlaid on the relative acceptance of an RNO station located at the South Pole
at $10^{17}$~eV.  The simulation assumes a resolution of the radio-frequency (RF) signal direction of $1^\circ$ \message{dzb: Is this one degree the incident neutrino direction resolution? If so, it's almost certainly way too optimistic. Or is it the resolution in the direction to the neutrino interaction point (which is of secondary importance, I think)}, polarization angle resolution of $10^\circ$, and off-cone Cherenkov angular resolution of $2^\circ$.
    }
    \label{fig:yearly_neutrinos}
\end{figure}

 Figures~\ref{fig:limits},~\ref{fig:yearly_neutrinos}, and Table~\ref{tab:event_rates} show the number of neutrinos expected from an IceCube $E^{-2.19}$ flux~\cite{8yearnumuflux} with the RNO baseline design outlined in this whitepaper.  If the astrophysical flux extends without a break, this discovery scale instrument will observe between 10 and 25 neutrinos in 5 years. 
 
 In Table~1, systematic
 uncertainties between independent simulations
 are observed at
 the level of a factor of 2.  The nature of these
 systematic differences 
 are being investigated. RNO will calibrate simulations
against RNO data where possible as has been done in
past arrays.

The discovery of the first neutrinos above $10^{16}$~eV will reveal the scale of detectors
 needed for the extremely high-energy 
 neutrino astrophysics program in the coming decades.
 A discovery will signal a hadronic mechanism
 for acceleration of cosmic rays in a new regime,
 and will probe cross sections at energies
 never before reached~\cite{Aartsen:2017kpd,Bustamante:2017xuy,Connolly:2011vc}, and even exceed previous constraints on Lorentz
 Invariance Violation~\cite{Gorham:2012qs,Aartsen:2017ibm} in the neutrino sector.
 A null result would indicate a break in the IceCube astrophysical flux, and constrain
 cosmogenic flux models derived from light
 composition and sources with a strong redshift dependence~\cite{Berezinsky:1975zz,BerezinskyZ,Stecker:1978ah,Hill:1983xs,Yoshida:1993pt,Engel:2001hd,Anchordoqui:2007fi,Takami:2007pp,Ahlers:2009rf,Aloisio:2015ega,Heinze:2015hhp,Romero-Wolf:2017xqe,AlvesBatista:2018zui,Moller:2018isk,vanVliet:2019nse}.

RNO's 
broad energy range enables studies of high-energy neutrino production mechanisms, as shown in Table~\ref{tab:event_rates} and Figure~\ref{fig:yearly_neutrinos}.  For example, RNO will be sensitive to models of transient bursts of neutrinos due to 
binary neutron star mergers~\cite{Metzger}, and test 
models of constant sources that may account for the UHE cosmic-ray flux
~\cite{FangPulsar}.
The low energy threshold of RNO allows searches for gamma-ray bursts (GRBs) with lower neutrino luminosity than have been previously conducted with radio neutrino experiments~\cite{ara_grb, anitaGRB}.

RNO will search for point sources of neutrinos within a broad declination band and with degree-scale pointing resolution. Figure~\ref{fig:yearly_neutrinos} shows the acceptance on the sky and the expected pointing resolution of RNO. Events observed by both the surface and deep components of each station will improve our angular reconstruction, which will enable even more sensitive multi-wavelength follow-up.

\begin{table}
    \centering
    \begin{tabular}{l c c||l c c}
\multicolumn{6}{c}{Expected Number of Neutrinos in 5 Years with a Full RNO}\\
\hline
Model & Lo & Hi  & Model & Lo & Hi\\  
\hline
IceCube Flux (no cutoff)~\cite{8yearnumuflux} & 9.8  &   25 &
Biehl, TDE~\cite{Biehl} & 1.8 &   4.9 \\ 
Boncioli, LLGRB~\cite{Boncioli} & 1.4 &  3.8 & 
Fang, NS-NS Merger~\cite{Metzger} & 11 &  29 \\ 
Fang, Pulsar (max)~\cite{FangPulsar} & 16 & 39 & 
Heinze, CR fit~\cite{Heinze} & 0.1 & 0.3 \\
Kotera, SFR1~\cite{Kotera2011} & 12  &  30 & 
Murase, AGN~\cite{MuraseAGN} & 10 &  25 \\ 
Van Vliet, 10\% proton~\cite{vanVliet} & 1.4 &   3.3 \\
\hline
\end{tabular}  \\[.5 ex]

    {
    \renewcommand*{\arraystretch}{1.4} 
    \begin{tabular}{|c|ccc|}
        \hline
        Cutoff on IceCube Flux & $10^{17}$ eV &   $10^{18}$ eV &  $10^{19}$ eV \\
        \hline
        Exp. No. of $\nu$'s (Hi) & 5.1 &  14 &  21 \\ 
        \hline
    \end{tabular}
     
    \caption{Number of neutrino events expected for the baseline RNO design (61 stations), for 5 years
    of integration and 100\% livetime.  We show the 
    bounds on the predicted event rates 
    as results from AraSim, the ARA
    detector simulation adapted for RNO (Lo),
    and the next-generation simulation framework that is underway (Hi).
    The bottom two rows show the number of expected neutrino events (using the new simulations) for an exponential cutoff to the measured IceCube flux at various energies~\cite{8yearnumuflux}.}
       \label{tab:event_rates}
    }
\end{table}

\subsubsection{Cosmogenic Neutrinos}
Five years of observations with RNO also has the potential to discover (if not reached by current instruments) 
cosmogenic neutrinos, and have unprecedented capability to measure of order ten events.
\message{dzb: This is true of ARA, no?  ac: yes, i shied 
away from language here at first due pressure in 
the past but I changed it now.}
UHE cosmic rays of sufficient energy undergo pion photo-production on the cosmic microwave background (CMB) as they propagate from their sources, creating a population of cosmogenic neutrinos~\cite{BerezinskyZ}. The flux and spectrum of such neutrinos is grounded in the UHE cosmic-ray mass composition measurements from the Telescope Array and Auger experiments, but is subject to model assumptions about the cosmological luminosity and chemical evolution of the sources. Figure~\ref{fig:limits} compares the sensitivity of RNO with several models of astrophysical and cosmogenic neutrino production using the next-gen simulations  
\footnote{PyRex (Next-gen): https://github.com/bhokansonfasig/pyrex, NuRadioMC (Next-gen): https://github.com/nu-radio/NuRadioMC,
  AraSim (Heritage):  https://github.com/ara-software/AraSim}.

Heinze~et~al.~\cite{Heinze}~assumes a single source model with a rigidity-dependent cutoff, 
consequently disfavoring protons at the highest energies and producing a small neutrino flux. van Vliet~et~al.~\cite{vanVliet} allows protons within the experimental limits of the cosmic-ray experiments, without making assumptions about sources. In the event of a non-detection, RNO will establish the world's best limits at the EeV energy scale, which will constrain source evolution and cosmic-ray mass composition.

\subsection{Event Rates}

The radio Cherenkov detection technique relies on coherent, impulsive radio signals that are emitted when any flavor of neutrino interacts in a dense material~\cite{ConnollyV},
due to a negative charge excess in the neutrino-induced cascade. If the wavelength of the radiation is longer than the transverse size of the shower, the resulting emission is coherent, leading to a large boost in total power for high-energy showers at frequencies $\lesssim1$~GHz.  
This so-called Askaryan effect~\cite{Askaryan} 
has been demonstrated in a series of accelerator-based experiments~\cite{sand,ice,salt,t510}.

Building an observatory for UHE neutrinos requires instrumenting enormous volumes of a radio-clear dielectric such as glacial ice.
The radio attenuation properties of ice have been directly measured at locations in Greenland and Antarctica~\cite{Barwick:2014rga,barrella,besson,ara_performance,avva1}, showing attenuation lengths in excess of 1~km at the South Pole~\cite{Barwick05}.

\section{  Extremely-High-Energy Neutrino Detection with RNO}
\label{sec:approach}

The RNO detector concept balances the desire for a high effective volume per station with the need to characterize neutrino event properties in this discovery instrument. The detector design is driven by five requirements:

\begin{itemize}
    \item[1.] Achieve sensitivity to neutrinos across a broad range of energies to target both astrophysical and cosmogenic neutrino fluxes, which is best achieved with low trigger thresholds.

    \item[2.] Achieve high livetimes critical for multi-messenger observations and improved sensitivity to diffuse neutrino fluxes. 

    \item[3.] Provide high-quality energy and direction reconstruction of each neutrino event for multi-messenger studies, which is best achieved by comprehensive observations with both a surface and a deep component of a single station.

    \item[4.] View a large fraction of the sky to enhance multi-messenger observations.
    \message{dzb: I would have thought that argues against putting a detector at South Pole, no?}
    \item[5.] Enhance discovery potential by achieving a high reconstruction efficiency of triggered neutrino events and minimizing backgrounds.
\end{itemize}

\begin{figure}
    \centering
    \includegraphics[width=0.4\textwidth]{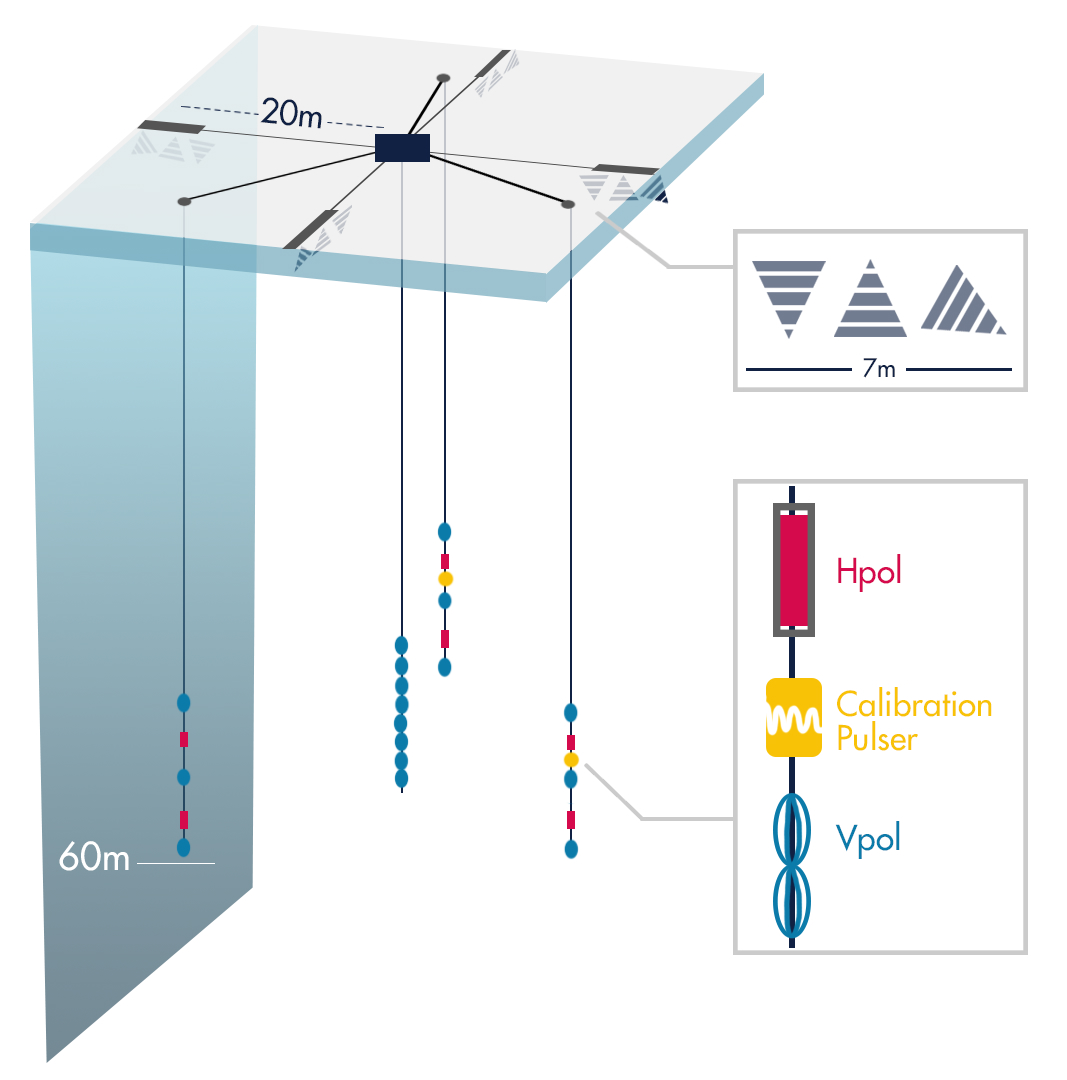}
    \includegraphics[width=0.5\textwidth]{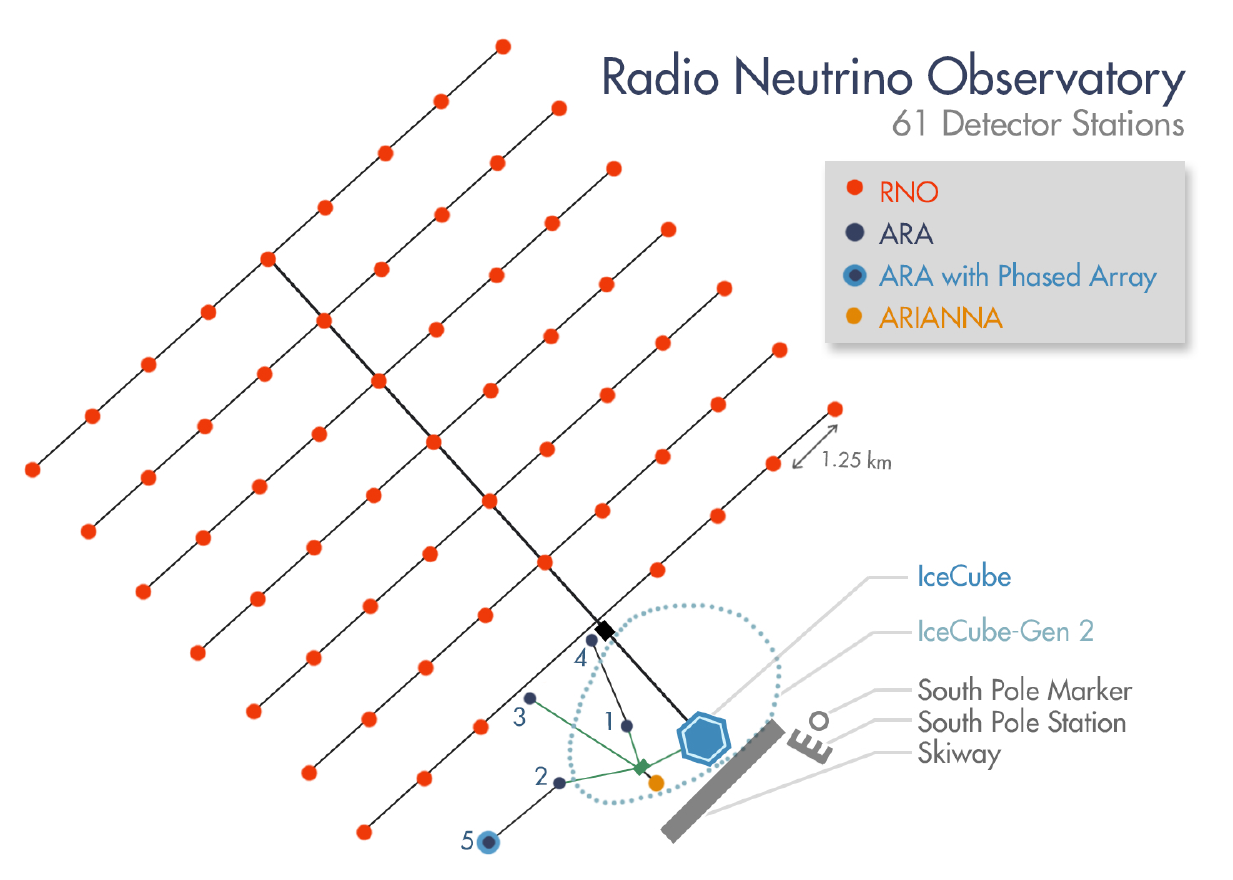}
    \caption{Left: A layout of one of the 61 stations in the proposed design.  A single station consists of two complementary parts: a surface array and a deep array.  The surface array is used for cosmic-ray detection and veto, background rejection, and neutrino event reconstruction. The deep array provides a large effective volume for neutrino detection per station.  The data acquisition system sits just under the surface, central to the station.  Right: The layout of the 61 stations, as viewed from above, relative to the existing ARA stations, ARIANNA test installation, and IceCube footprint at the South Pole.  The power and communications grid is also shown.
}
    \label{fig:RNOstation}
\end{figure}

Detailed simulation studies lead to the RNO design which consists of an
array of 61 stations with four strings of antennas each to 60~m depth with an additional array of antennas at the surface, shown in Figure~\ref{fig:RNOstation}. The deep component consists of a 50-m-centered, 8-channel central interferometric phased array trigger string and three reconstruction strings spaced 20~m apart with a total of 15 dedicated antennas (both horizontally and vertically polarized) that are used for event reconstruction. The surface component consists of 12 high-gain log-periodic dipole antennas (LPDAs) placed in trenched slots in the snow. 
Stations are spaced along a grid with a spacing of 1.25~km, to allow the stations to observe nearly independent volumes of ice.

\subsection{Low Thresholds with an Interferometric Phased Array Trigger}

We will use an interferometric phased array to trigger RNO, which has been successfully demonstrated {\it in situ} at the South Pole on the ARA experiment~\cite{oberla, vieregg} and has achieved the lowest demonstrated trigger threshold in the field.  Since the astrophysical neutrino flux is a falling spectrum, being able to see lower energy events dramatically increases the neutrino event rate observed.  Pushing the trigger threshold down by using this full-waveform phased array trigger technique allows RNO to have a lower energy threshold and better overlap with IceCube.

The phased array trigger coherently sums the full radio waveforms with time delays corresponding to a range 
of angles of incident plane waves, giving a boost in 
signal-to-noise ratio (SNR) for triggering that goes as $\sqrt{N}$, where $N$ is the number of 
antennas in the array.  
With different sets of delays, we create multiple effective antenna beam patterns that together
cover the same solid angle as each individual antenna but with higher gain.
The improvement in SNR for triggering directly translates into a lower energy threshold for finding neutrinos.
Projecting the performance of the existing ARA phased array system~\cite{oberla} to an 8-channel system, we expect to achieve an elevation-averaged 50\% trigger efficiency point of 1.5$\sigma$.  

\subsection{High Analysis Efficiencies and Low Backgrounds}
\label{sec:background}

In addition to triggering on and extracting event parameters from neutrino events, we must be able to separate any neutrino events in our recorded data set from the three major sources of backgrounds: incoherent thermal noise, impulsive man-made noise, and radio impulses resulting from cosmic-ray air showers.  
RNO builds on a variety of instruments that have been developed in the radio detection community to ensure event purity.

The RNO instrument is designed to reject man-made and cosmic-ray backgrounds both by triggering from deep within the ice (where the backgrounds are low compared to at the surface) and by incorporating a surface veto. ARA has shown that the man-made and thermal backgrounds decrease for receivers deployed deeper in the ice~\cite{ara_testbed}, achieving a background on the most recent analysis of 0.01 background events in each of two stations in ten months of data~\cite{ara_2station}. 
Perhaps the most significant recent advance was the proven ability for a deep receiver array to accumulate high-quality data during the austral summer months, during the peak of South Pole station activity, the prerequisite for a high livetime.
With the surface array component of RNO, we will additionally be able to identify and characterize in-ice signals from cosmic-ray showers~\cite{deVries}. With the deep stations and the enhanced veto offered by the surface array, we expect the backgrounds to be $\le$0.01 per station per year.

\section{Technical Approach: The RNO Design}

\begin{figure}[ht]
    \centering
    \includegraphics[width=5in]{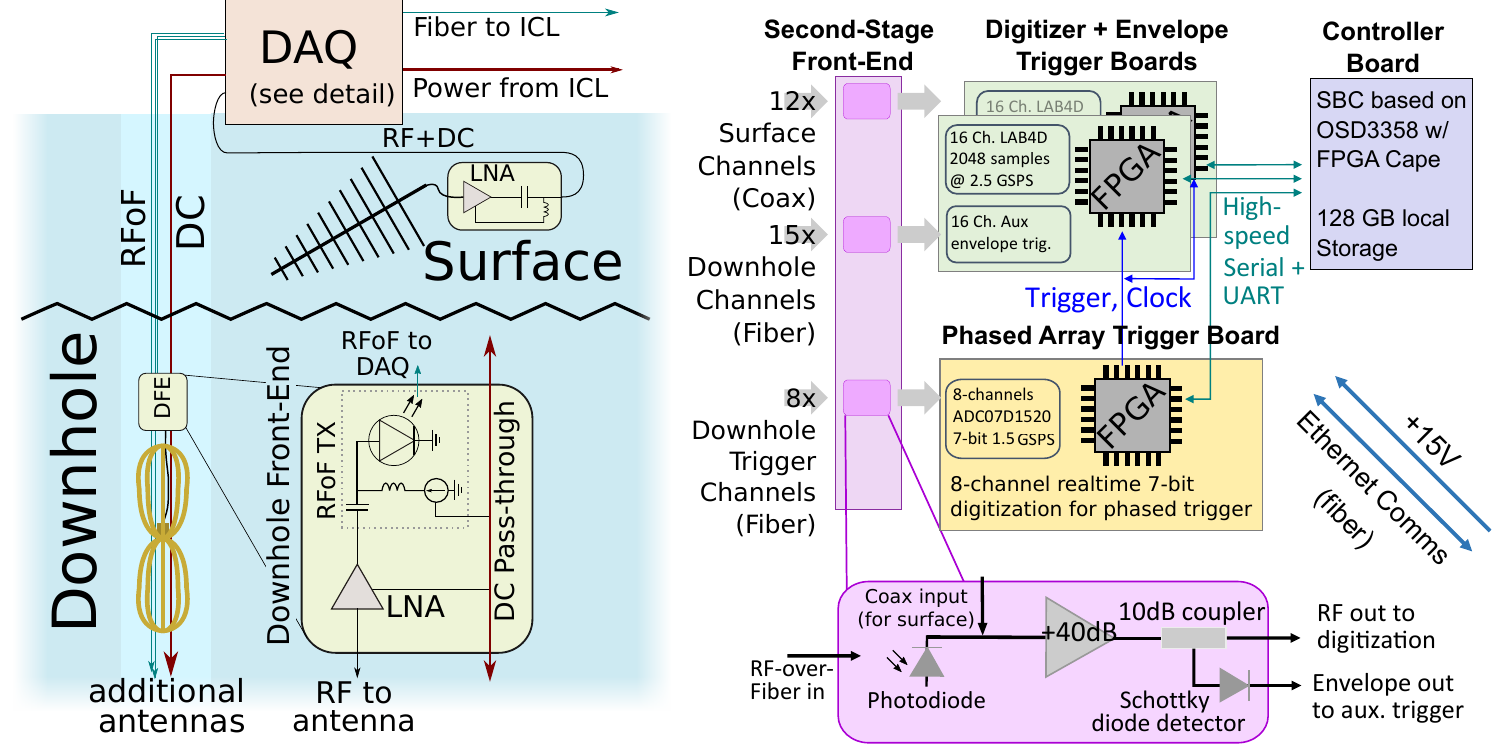}
    \caption{Left: RNO Station schematic, showing both the surface and deep components.  
        Right: RNO DAQ system. The deep component uses a phased array trigger; the surface array, a diode-based trigger. Signals are recorded with LAB4D-based digitizer boards. A control board communicates with both the trigger and digitizer boards.  
        }
    \label{fig:blockdiagram}
\end{figure}

{\bf Radio Frequency Front-End Design:} 
As successfully demonstrated with ARA, the deep stations will use vertically-polarized bicone antennas and horizontally-polarized ferrite-loaded quad-slot antennas for the deep component front-end. 
To minimize the noise temperature of the system, each antenna feed is connected with a short coaxial cable to a downhole front-end, where a low-noise amplifier (LNA) boosts the signal strength. To prevent a significant gain slope from losses in a long length of coaxial cable, each front-end contains a Radio Frequency over Fiber (RFoF) transmitter. The RFoF link and LNA are both powered by a DC connection from the surface, which will be the only through-going coaxial cable in the array. 

At the surface, four upward-pointing and four downward-pointing antennas measure
the two components of horizontal polarization well, and provide a clear up-to-down ratio of signal strength to distinguish upgoing versus downgoing signals. The vertical polarization component is captured with angled LPDAs, as shown in Figure~\ref{fig:RNOstation}. 

\label{sec:daq}

{\bf Triggering, Digitization, and Data Acquisition}
 The DAQ will accommodate up to 16 surface channels and 24 downhole channels with the ability to sustain a station event-rate of 10 Hz.
The primary trigger will be a phased-array system, closely based on the NuPhase system deployed at  ARA5~\cite{oberla}. The NuPhase trigger is built on the streaming digitization of a compact array of 8 vertically-polarized antennas at 7-bit resolution and 1.5 GSPS. A power target of \textless 20\,W for this board is achievable. The beamforming trigger will be formed on an FPGA and provide a low-threshold trigger that will be broadcast to the digitization boards. The event waveforms from the 8-channel NuPhase array will be sent to the control board to be packetized with the LAB4D digitized channels.

{\bf Power and Communications}
In an environment with power infrastructure such as the South Pole, 
the power and communications grid was determined to be
a cost effective solution that ensures year round power and communications at a small fraction ($<7$\%) of the total project cost.  A conceptual design has been developed.
For the South Pole a DC distribution grid (see Figure~\ref{fig:RNOstation}) is realistic.  

However, 
as part of RNO's role as pathfinder for the radio
component of IceCube-Gen2,
we will develop 
alternative power solutions such as solar panels and wind turbines,
and optimize
electronics for low power. 
We anticipate that an array of the size envisioned for IceCube-Gen2 will need to rely on autonomous power.

If Greenland is the preferred site, autonomous power would be used, and conceptual designs for autonomous power exist. 
Summit station in Greenland offers significantly less logistical support capabilities than the South Pole.
Greenland's stronger winds compared to at Pole, and shorter periods of darkness, make it more
amenable to solar and wind power.
  Data transfer would occur via Iridium, which can be powered with only a few Watts per station.

{\bf Deployment of Instrumentation}
The deep holes are drilled using an Auger with add-in drill sections (ASIG) from the University of Wisconsin's Ice Drilling Program (IDP).  
The cost for drilling is at the 1\% level of the project cost and the personnel required for drilling is 3 to 5 people on the ice,  depending on the number of holes to be drilled.
The deployment of the instruments is relatively straightforward
based on prior field experience both at the South Pole and in Greenland.

{\bf  Detector Calibration Requirements and Strategies}
The calibration of RNO is critical to the interpretation of the data and, ultimately, our ability to claim the observation of a signal.  The requirements for calibration are two-fold.  First, we must measure the station geometry and signal chain performance {\it in situ}. 

Calibration of the RNO detector array is comprised of three primary methods:
\begin{itemize}
\item[1.] Pre-deployment calibration and characterization in the lab. 
\item[2.] `Local' intra-station calibration of the antenna locations and signal chain at the station level.
\item[3.] `Global' inter-station calibration of receivers at the observatory-wide level, using distant, bright transmitters located a few kilometers away.
\end{itemize}

Two calibration devices will be located at each station, deployed in the same holes as the station antennas.  The devices will be similar to the ones used by ARA, and are capable of transmitting a sub-nanosecond pulse or noise with a flat power spectrum over the sensitive frequency range.

{\bf Timeline and Cost}
The anticipated timeline of RNO is shown
in Fig.~\ref{fig:yearly_neutrinos}.  
The year of the 
first deployment will depend on the timeline for
availability of resources and logistics at the chosen 
site.  The cost of the project falls under the Small
category for ground-based projects.  We anticipate the 
lifespan of the project to continue until the size
of RNO is overtaken by the radio component of  IceCube-Gen2.

\section{Summary}
\label{sec:summary}

The Radio Neutrino Observatory (RNO) is
the mid-scale, discovery phase, extremely 
high-energy neutrino telescope that will
probe the astrophysical neutrino flux at
energies beyond the reach of IceCube.
The $\sim 20$ RNO stations deployed in the first two 
 years will serve as a pathfinder for 
 the radio component of IceCube-Gen2, informing the design of stations that are fully optimized for delivering 
 the most science for the cost, in
 the first year of IceCube-Gen2 deployments.

RNO is designed for a diverse science program. This includes
multi-wavelength neutrino observations when coupled
with IceCube measurements, multi-messenger astronomy
through alerts and targeted searches for source coincidences, and constraining the nature of the 
highest energy accelerators through measurements spanning four decades in energy.

Having RNO in operation contemporaneously
with
IceCube's continued characterization of
the astrophysical flux will be an important
opportunity for multi-wavelength observations
of the same sources.  With both RNO and IceCube
at the South Pole, the portion of the sky that is viewable by both projects is a band within a few degrees of the horizon.  If RNO is deployed
in Greenland, then there would be larger overlap in sky coverage, and RNO would sweep more sources at lower latitude although not continuously.  Furthermore, in this case IceCube events at lower energies could be used because of the Earth shielding effect.

The RNO collaboration brings together 
broad expertise from the worldwide community in the field of radio detection of high-energy particles. 
The collaboration includes all members of the ARA collaboration, 
several members of the ARIANNA
collaboration, and new groups that have joined the RNO collaboration without prior history on either project. 
With the RNO collaboration, a truly international consortium has formed 
that incorporates substantial contributions from international partners.  Coordination between
the RNO and IceCube-Gen2 collaborations with regard
to the intersection of timelines, logistics, and synergies of
the two projects is in progress.

\section{Acknowledgements}
We are grateful for
Belgian Funds for Scientific Research (FRS-FNRS and FWO) and the FWO programme for
 International Research Infrastructure (IRI).

\newpage
\bibliographystyle{custom_style}
\bibliography{main}

\end{document}